\documentclass[conference, 10pt]{IEEEtran}
\IEEEoverridecommandlockouts

\def\BibTeX{{\rm B\kern-.05em{\sc i\kern-.025em b}\kern-.08em
    T\kern-.1667em\lower.7ex\hbox{E}\kern-.125emX}}

\usepackage{cite}
\usepackage{amsmath,amssymb,amsfonts}
\usepackage{algorithmic}
\usepackage{graphicx}
\usepackage{textcomp}
\usepackage{xcolor}
\usepackage{comment}
\usepackage{footnote}
\usepackage{hyperref}
\usepackage{gensymb}
\usepackage{multirow}
\usepackage{subcaption}

\begin{document}

\title{Impact of Data Compression on  Downstream AI Tasks: A Study  using Teleoperated Driving over 5G
}

\author{
\IEEEauthorblockN{Qixin Zhang, Steven Sleder, Xinyue Hu, Faaiq Bilal, Wei Ye, Zhi-Li Zhang}
\IEEEauthorblockA{University of Minnesota Twin Cities, Minneapolis, USA \\
    \{zhan8548, slede001, hu000007, bilal021, ye000094\}@umn.edu \qquad zhzhang@cs.umn.edu}
}

\maketitle

\begin{abstract}
Teleoperation, such as remote driving, is considered as a key use case of  5G and Next-Generation (NextG) networks. In this context, robots, autonomous vehicles, or other autonomous agents transmit sensor data over mobile networks to edge or cloud servers, where AI systems collaborate with human operators to provide situational awareness and enable remote control. In the case of teleoperated driving,  
vehicles are equipped with an array of cameras and LiDAR devices, which can generate 100s Mbps (megabits per second) of data. As shown in existing measurement studies, such data volumes far exceed the \emph{uplink} capacity of currently deployed 5G networks, especially when multiple vehicles compete for radio resources. Data compression is thus imperative. In this paper, we explore the impact of sensor data compression on the performance of downstream AI tasks running in edge/cloud servers, which are crucial to alert human operators for safe teleoperation.  Using object recognition and semantic segmentation as two example AI tasks,
we study how data compression affects the performance of these two AI tasks using unimodal (video or LiDAR) and multi-modal (video+LiDAR) data. We find that lossy data compression generally decreases the performance of AI tasks. 
The performances of these AI tasks exhibit differing degrees of sensitivity based on the types of data sources and levels of compression. We also empirically identify an optimal trade-off point for the multi-modal vision tasks.


\end{abstract}

\section{Introduction}

Teleoperation of robots, drones, and vehicles is considered as a key use case for 5G and next-generation (NextG) wireless networks, and have a wide range of applications~\cite{app-1,app-2,app-3}.
For example, today's autonomous vehicles can at best achieve Level-4 autonomy~\cite{SAE-level}, which means that they can drive autonomously only under certain \emph{operational design domain} (ODD). When encountering complex scenarios outside the ODD, human intervention is required. This is where teleoepration comes into play: a human operator working in a remote teleoperation center near the edge/cloud facility can take over the control of a vehicle encountering a scenario (e.g., a work zone) outside its ODD and remotely navigate the vehicle through such a scenario. In order to safely operate the vehicle remotely, delivering critical sensor data such as camera and LiDR data over 5G/NextG networks in a timely manner is critical to provide \emph{situational awareness}. Downstream  AI tasks such as object detection and segmentation running on edge/cloud servers to generate alerts and early warnings are also needed in assisting human operators in teleoperation.


Today's autonomous vehicles are often equipped with a multitude of sensors such as radars, cameras, and LiDAR devices.
For instance, each Baidu Apollo vehicle~\cite{Apollo} has 13 cameras, 5 millimeter-wave radars, and 2 LiDAR sensors, while a Waymo vehicle~\cite{Waymo-Driver} features a total of 29 cameras, 6 radars, and 4 LiDAR sensors. The high-precision sensors on a single vehicle can generate data volumes of 100s of Mbits of data (see \textbf{Sec \ref{sec:sensor_requirement}} for more details). Due to limited viewpoints from a single vehicle,  sensor data from nearby vehicles or from fixed camera/LiDAR sensors from the road infrastructure (e.g., at a busy intersection) will also need to be delivered and fused for safe teleoperation in complex environments. All these pose significant bandwidth and latency burdens on the underlying mobile networks.



\begin{figure}[t]
    \begin{subfigure}[t]{0.49\textwidth}
        \centering
        \includegraphics[width=0.485\linewidth, height=1.2in]{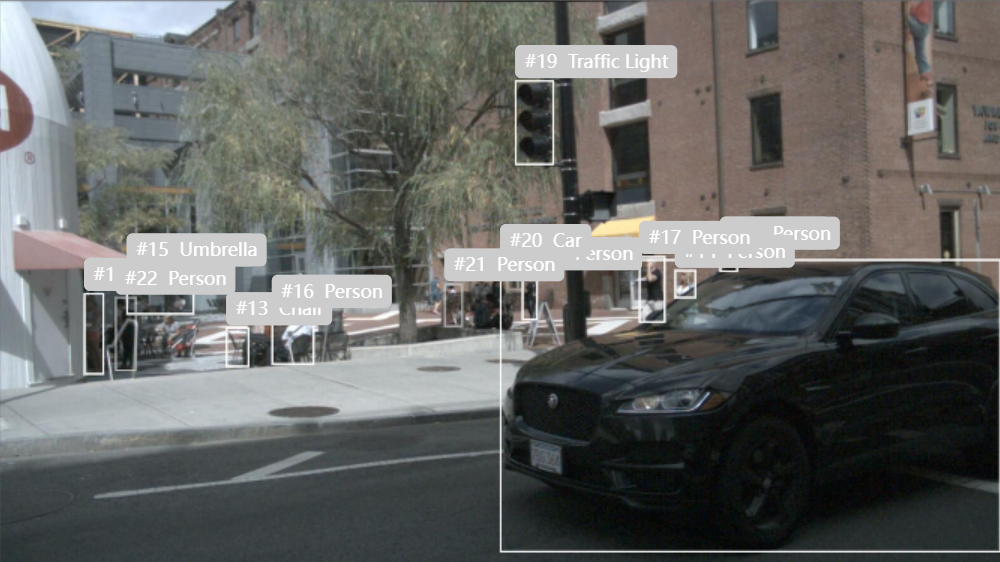}
        \includegraphics[width=0.485\linewidth, height=1.2in]{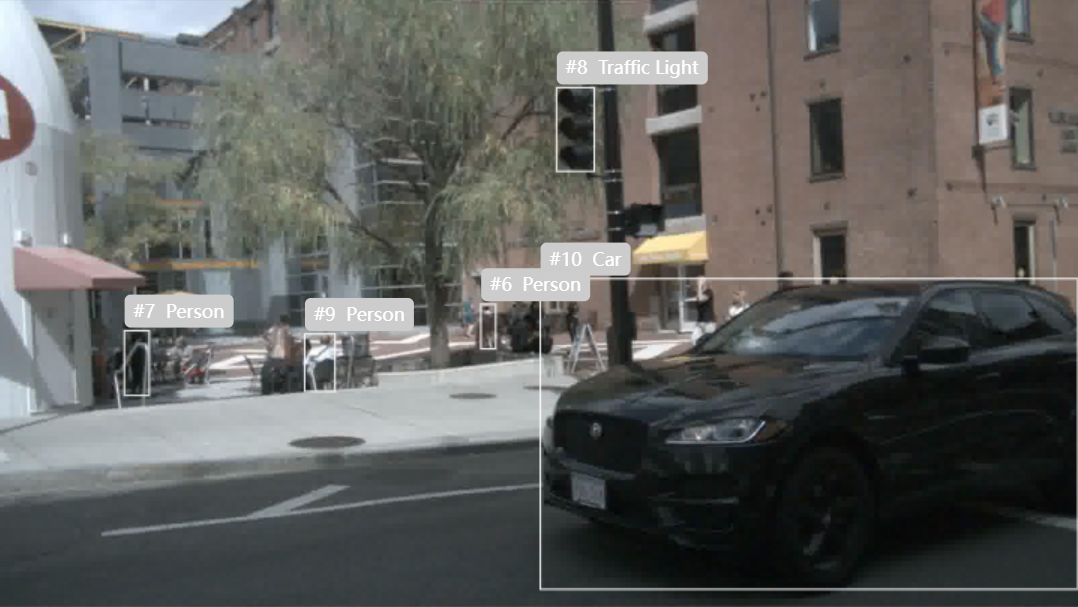}
        \caption{Camera image data.}
        \vspace{0.25em}
    \end{subfigure}
     ~
    \begin{subfigure}[t]{0.49\textwidth}
        \centering
        \includegraphics[width=0.485\linewidth, height=1.2in]{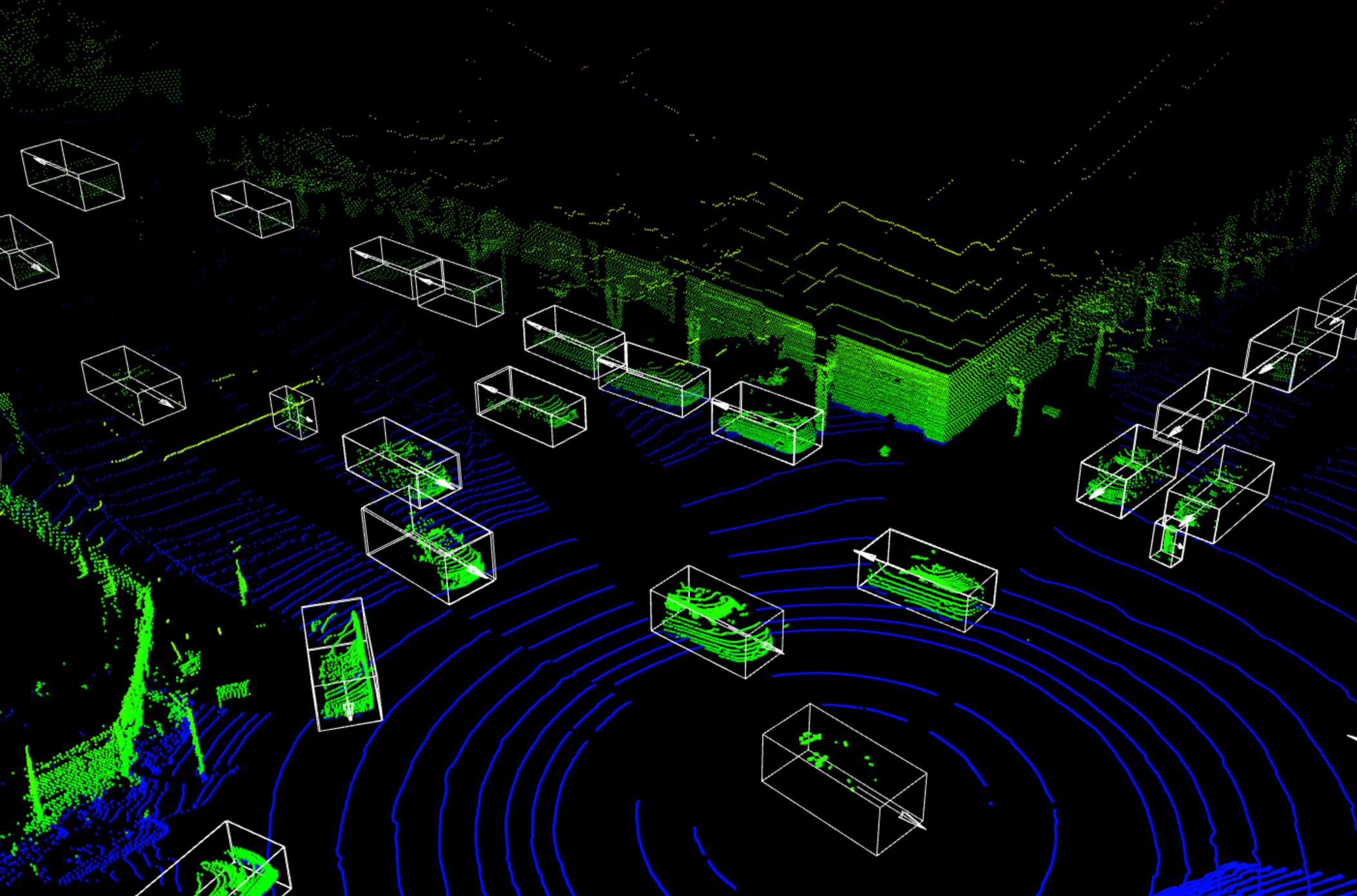}
        \includegraphics[width=0.485\linewidth, height=1.2in]{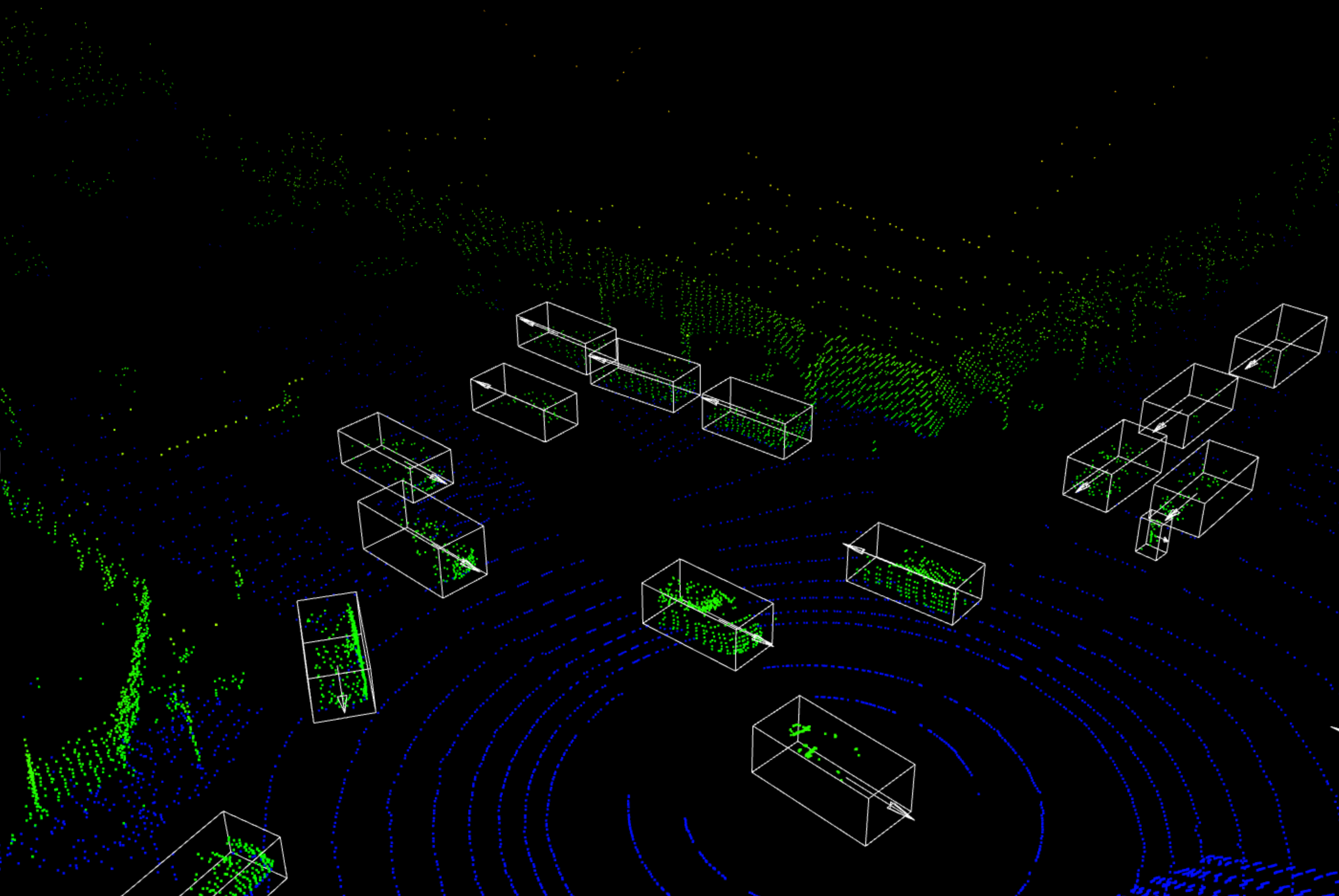}
        \caption{LiDAR point clouds.}
    \end{subfigure}
    \caption{Visualization results of 3D object detection on image and point cloud data. Compared to raw data (left column), AI detects fewer objects in the compressed data (right column).}
    \label{fig:vis_results}
    \vspace{-1.5em}
\end{figure}

While 5G networks provide significantly higher bandwidth over 4G networks, the improvements largely manifest in the downlink performance~\cite{carpenter23data, ye2023closer, rochman2023comprehensive}. In other words, there is a notable \emph{asymmetry} in uplink and downlink: uplink throughput is significantly lower than downlink throughput and quite far from meeting the bandwidth requirements (of "raw" sensor data) from a single vehicle (see \textbf{Sec \ref{sec:5g_up_bottleneck}}), not to mention the bandwidth requirements for delivering sensor data from multiple vehicles. Therefore, compressing sensor data such as video and LiDAR before transmission is imperative in reducing the uplink bandwidth requirements.  However, compressing the sensor data too aggressively may lead to the distortion of important information in the sensor data, thereby affecting the visual perception of human operators or the efficacy of the downstream AI tasks in object detection, recognition, and tracking. This may potentially lead the human teleoperator to make erroneous decisions, therefore posing safety risks. Understanding the trade-off in bandwidth reduction by compressing sensor data for delivery over mobile networks and its impact on the efficacy of downstream AI tasks is therefore an important question that has not gained adequate attention in the research community. 

This paper is devoted to exploring this important question. 
Using teleoperated driving as a key use case, our goal is two-fold: we i) not only aim to  \textit{quantify} how data compression affects the performance of downstream AI tasks by considering two AI tasks, object detection and semantic segmentation -- both of which are crucial to teleoperation, using unimodal (camera or LiDAR) and multimodal (camera and LiDAR) sensor data; ii)  but also attempt to empirical \textit{identify} the optimal trade-off points in data compression and downstream AI performance for these AI tasks and uni/multimodal data types.
To this end, we utilize RGB images collected from the onboard cameras and LiDAR point clouds, compressing each at different ratios. We employ the state-of-the-art vision model, BEVFusion, as a benchmark, which can perceive the surrounding environment using either single-modal (i.e., using only one type of sensor data) or multi-modal (i.e., using both sensor data) inputs. We evaluate two key downstream tasks: object detection and semantic segmentation. \textbf{Sec \ref{sec:problem_statement_Methodology}} provides more details about our methodology.

Our experimental results in \textbf{Sec \ref{sec:result}} reveal that 
lossy data compression generally decreases downstream AI performance, exhibiting a non-linear and non-uniform relationship between the level of compression and algorithmic effectiveness. LiDAR and camera information excel in different vision tasks: LiDAR outperforms in object detection, while cameras surpass in segmentation, with each detecting certain objects more effectively. Furthermore, we identify the optimal compression trade-off for multi-modal vision tasks that utilize both the LiDAR and camera information.

\section{Background \& Related Works}


In this section, we introduce performance metrics used to characterize 5G network performance. We also include a review of related work on related AI tasks performed at a teleoperation station. In addition, we write about compression techniques used for video and Lidar data and their impact on the performance of downstream AI tasks.

\begin{figure}[t]
\begin{minipage}{.95\linewidth}
    \centering
    \vspace{-4mm}
    \includegraphics[width=0.7 \textwidth]{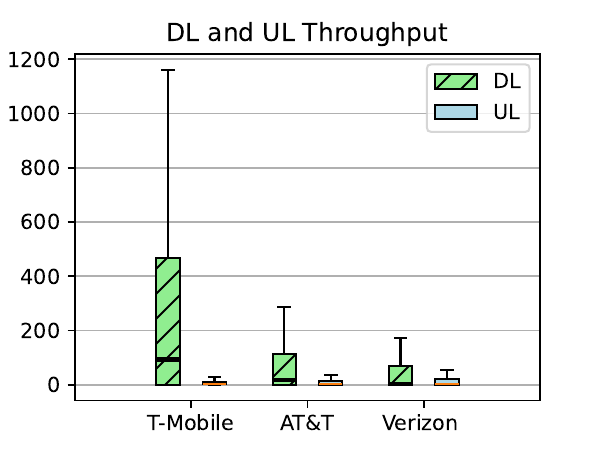}
    \vspace{-2mm}
    \caption{The variation and asymmetry in Downlink and Uplink Throughput for different carriers in the US}
    \vspace{-4mm}
    \label{fig:5g_asymmetry}
\end{minipage}
\end{figure}

\begin{figure}[t]
\begin{minipage}{.95\linewidth}
    \centering
    \includegraphics[width=0.7 \textwidth]{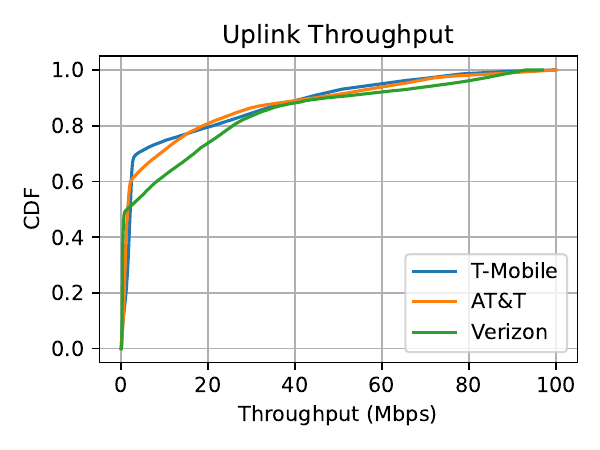}
    \vspace{-2mm}
    \caption{CDF of Uplink Throughput in 5G Networks with different carriers in the US.}
    \vspace{-4mm}
    \label{fig:5g_uplink_cdf}
\end{minipage}
\end{figure}

\subsection{Requirement for sensory data transmission} 
\label{sec:sensor_requirement}
Recent work documents the throughput requirements of transmitting Lidar and video data over a network. \cite{carpenter23data} The paper reports a requirement of 45.6 Megabits per second (Mbps) for a 64-beam Lidar and 2.73 Mbps for a single camera. Various sources, including the nuScenes dataset (detailed in Section \ref{sec:data&bench}) report the use of the collection of sensors for a teleoperated vehicle. \cite{caesar2020nuscenes} Based on our calculation from information available via nuScenes, a 32-beam Lidar feed will require a throughput of 44.48 Mbps. As a result, for a single vehicle, the throughput requirements can be several hundreds of Mbps. We make use of this information along with our throughput measurements over cellular to assess whether cellular uplink is capable of supporting uncompressed data transmission for teleoperated vehicles.  

\subsection{5G Uplink Performance as a Bottleneck}
\label{sec:5g_up_bottleneck}
We conducted and made use of real-world 5G downlink and uplink throughput measurements to evaluate the need for data compression. Figure \ref{fig:5g_asymmetry} shows a summary of the uplink and downlink performance of commercial 5G networks deployed by different operators. \textit{Uplink throughput} refers to the rate at which data is transmitted from the user's device to a remote server on the internet (in downlink, the direction is reversed). We observe a large asymmetry, ranging from 8$\times$ to 15$\times$ in the uplink and downlink throughput, which occurs because cellular networks are optimized for downlink-heavy smartphone use. This discrepancy is a result of the use of throughput boosting techniques such as Multiple-In-Multiple-Out (MIMO) and Carrier Aggregation (CA) in downlink, while they are absent in uplink and the allocation of fewer slots (time) for uplink compared to downlink. \cite{carpenter23data}  

Additionally, Figure \ref{fig:5g_uplink_cdf} shows that uplink throughput stays under 40 Mbps for 90\%, under 20 Mbps for 80\%, and under 10 Mbps for almost 70\% of our measurements. This behavior is consistent across carriers. This information, coupled with the throughput requirements discussed in \ref{sec:sensor_requirement} make the basis of our argument that cellular networks cannot support the upload of uncompressed LiDAR and camera data.

\subsection{Impact of Compression Techniques}
The impact of data compression techniques on downstream AI performance has been studied in \cite{hase2011influence, poyser2021impact, bhowmik2022lost, chan2023influence, sakthi2024impact}. 
However, these studies are limited in scope and diversity, focusing mainly on image or video data compression for specific environments (e.g., 2D~\cite{poyser2021impact}, nighttime~\cite{hase2011influence}) or objects (e.g., vehicles~\cite{chan2023influence}, pedestrians~\cite{hase2011influence}) using only camera-source information. For example, 
Poyser et al.~\cite{poyser2021impact} analyze the effects of common image and video compression techniques on CNN-based video analytics in 2D environments.
Bhowmik et all.~\cite{bhowmik2022lost} assess how lossy JPEG compression impacts infrared imagery for CNN-based object detection. 
Sakthi et al.~\cite{sakthi2024impact} investigate the influence of standard video compression on fisheye camera videos for 3D object detection in autonomous driving systems.

To the best of our knowledge, we are the first to examine the compression of multi-modal data sources, incorporating both camera and LiDAR data. We quantify the impact on object detection and segmentation using advanced deep learning architectures (Vision Transformer and CNN) in 3D environments for various object types. Additionally, we explore how unimodal and multimodal visual perception tasks exhibit differing sensitivities to data compression.

\section{Problem Statement \& Methodology}
\label{sec:problem_statement_Methodology}

In this section, we outline the research question, methodology, dataset, and evaluation model used in our experiments.

\subsection{Problem Statement}

As discussed earlier in \ref{sec:5g_up_bottleneck}, 5G Uplink throughput cannot support LiDAR and video data transfer to an edge or cloud server. Hence, we require data compression to support its transmission. However, data compression has an impact on the performance of AI tasks at the edge or cloud. Hence, our goals for this paper are: 
\begin{enumerate}
    \item Quantify the performance impact of data compression on AI tasks
    \item Understand and quantify the trade-off in different compression strategies for video and LiDAR
\end{enumerate}



\subsection{Dataset and Benchmarks}
\label{sec:data&bench}
This subsection provides the details of our dataset \textit{nuScenes} and the evaluation model \textit{BEVFusion}. \cite{caesar2020nuscenes ,liu2022bevfusion}

Our evaluation model, BEVFusion, is pre-trained on the full nuScenes dataset~\cite{caesar2020nuscenes}, a large-scale outdoor dataset used for autonomous driving, containing diverse annotations to support various tasks, such as 3D object detection, tracking, and BEV map segmentation for 1000 samples with a train-validate-test split of 700-150-150. Each annotated sample in the nuScenes dataset includes six monocular camera images with a 360-degree field of view and a 32-beam LiDAR scan.

Meanwhile, as an initial exploratory work, our experiments are performed on a subset of the full nuScenes dataset, referred to as nuScenes-mini, consisting of 404 samples taken from only the training and validation sets of the complete dataset. We utilize only the 81 samples of the nuScenes-mini validation set in our experiments as it provides ground truth labels which allow us to quantify error without biasing the performance of the model to data on which it was explicitly trained. This setup ensures that our evaluation is both rigorous and reflective of real-world performance, allowing us to draw meaningful conclusions about the impact of data compression on 3D object detection.

\subsection{Compression Method}
We vary the resolutions of both the RGB camera images and LiDAR point clouds for comparison on single- and multi-modal tasks in an exhaustive manner. Images are downsampled by varying the quality parameter of the Motion JPEG codec's intraframe compression. We select the quality parameter from the range of [5, 15, 20, 25, 30], with higher values resulting in lower image quality and, trivially, resulting in a decrease in image size. For example, an image with a quality of 5 remains visually similar to the original while resulting in a x5 reduction in data size.

Simultaneously, the resolution of the LiDAR point clouds produced is downsampled using simple voxelization, creating downsampled sets resulting from voxel sizes in the range of [0.1, 0.25, 0.5, 0.75, 1.0]m$^3$. Larger voxel sizes result in the contraction of points within a larger area, likewise resulting in lower resolution and a reduction in data size. The point clouds are downsampled individually and as with the intraframe compression of MJPEG, there is no temporal relationship in the reduction of their quality.

\subsection{Downstream Vision Tasks in 3D Environments} 

\subsubsection{Object Detection}
Object detection identifies and locates objects within a 3D environment based on their shape, position, and orientation. It involves detecting objects' presence and determining their spatial coordinates in real time. This task is vital for applications such as autonomous vehicles, robotics, and augmented reality, where accurate and timely object recognition and spatial understanding are essential for navigation, interaction, and safety. BEVFusion~\cite{liu2022bevfusion} is a multi-task multi-sensor fusion framework that fuses both camera and LiDAR information for 3D perception tasks. For 3D object detection, BEVFusion utilizes Swin-Transformers~\cite{liu2021swin} for image feature extraction, VoxelNet~\cite{zhou2018voxelnet} for LiDAR feature extraction and a convolution-based Bird's-Eye View (BEV) encoder to merge these features. The detection head includes a center heatmap and regression heads to predict object locations, sizes, rotations, and velocities. We evaluate the mean Average Precision (mAP) of BEVFusion for object detection across various compression levels of camera and LiDAR data.

\subsubsection{Semantic Segmentation}
Semantic segmentation partitions a 3D point cloud or image pixels into semantically meaningful parts or regions. The objective is to identify and label various objects and components within a 3D scene. This process is crucial for applications such as robotics, autonomous driving, and augmented reality, where accurate recognition and labeling of 3D elements enhance interaction, navigation, and environmental understanding. 

For 3D semantic segmentation, BEVFusion utilizes the same extracted features used for object detection but employs a separate, independent segmentation head. This head performs binary segmentation for each map category using focal loss. Using the source code and pre-trained models of BEVFusion, we evaluate the \textit{mean intersection over union} (mIoU) of segmentation across various compression levels of camera and LiDAR data. 

\section{Experimental Results}
\label{sec:result}

In this section, we quantify the impact of various levels of data compression on the pre-trained BEVFusion under its respective evaluation metrics. We summarize our findings in the following:
\begin{itemize}
    \item Lossy compression decreases downstream AI performance, exhibiting a non-linear and non-uniform relationship between the level of compression applied and algorithmic performance.
    \item LiDAR and camera information excel in different vision tasks: LiDAR outperforms in object detection, while cameras surpass in segmentation, with each detecting certain objects better. 
    \item We find the optimal compression trade-off in multi-modal vision tasks.
\end{itemize}

\begin{figure*}[t] 
\begin{minipage}{.3\linewidth}
    \centering
    \includegraphics[width=\textwidth, height=1.7in]{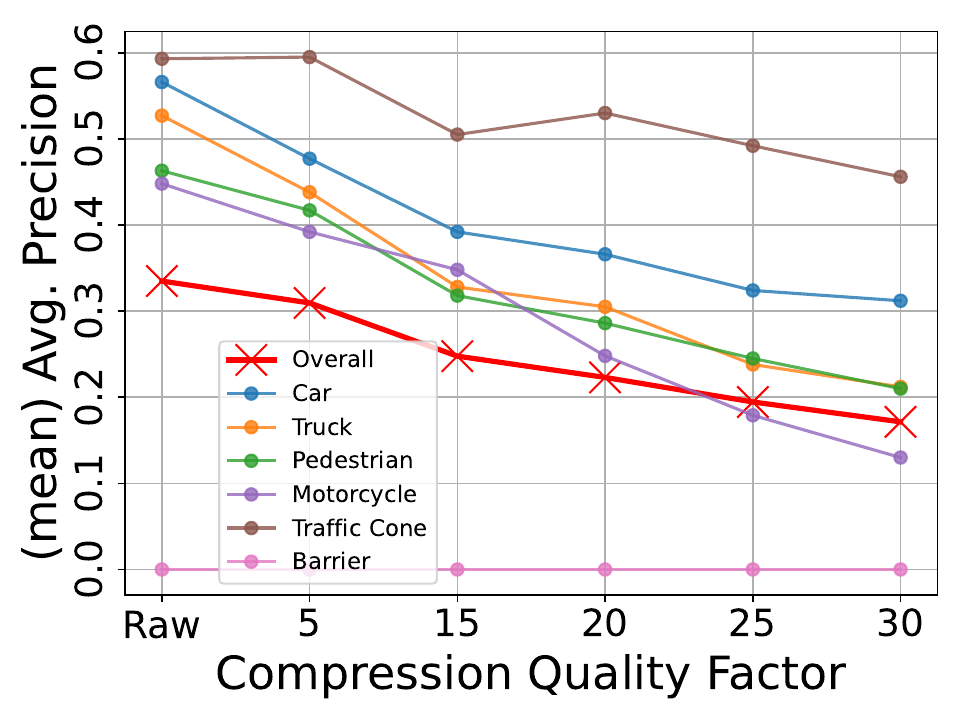} 
    \caption{Impact of compression quality factor on mAP of various object categories. }
    \label{fig:compression_quality}
\end{minipage}
\hfill
\begin{minipage}{.3\linewidth}
    \includegraphics[width=\textwidth, height=1.7in]{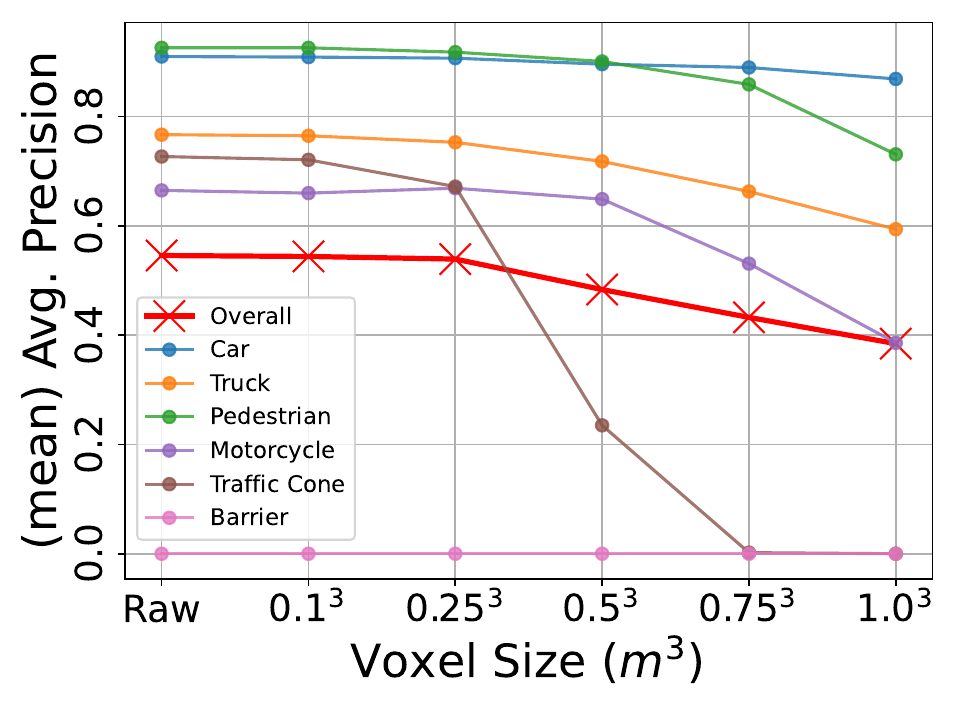}
    \caption{Impact of voxel size on mAP of various object categories. }
    \label{fig:voxel_size}
\end{minipage}
\hfill
\begin{minipage}{.37\linewidth}
    \centering
    \includegraphics[width=0.85\textwidth]{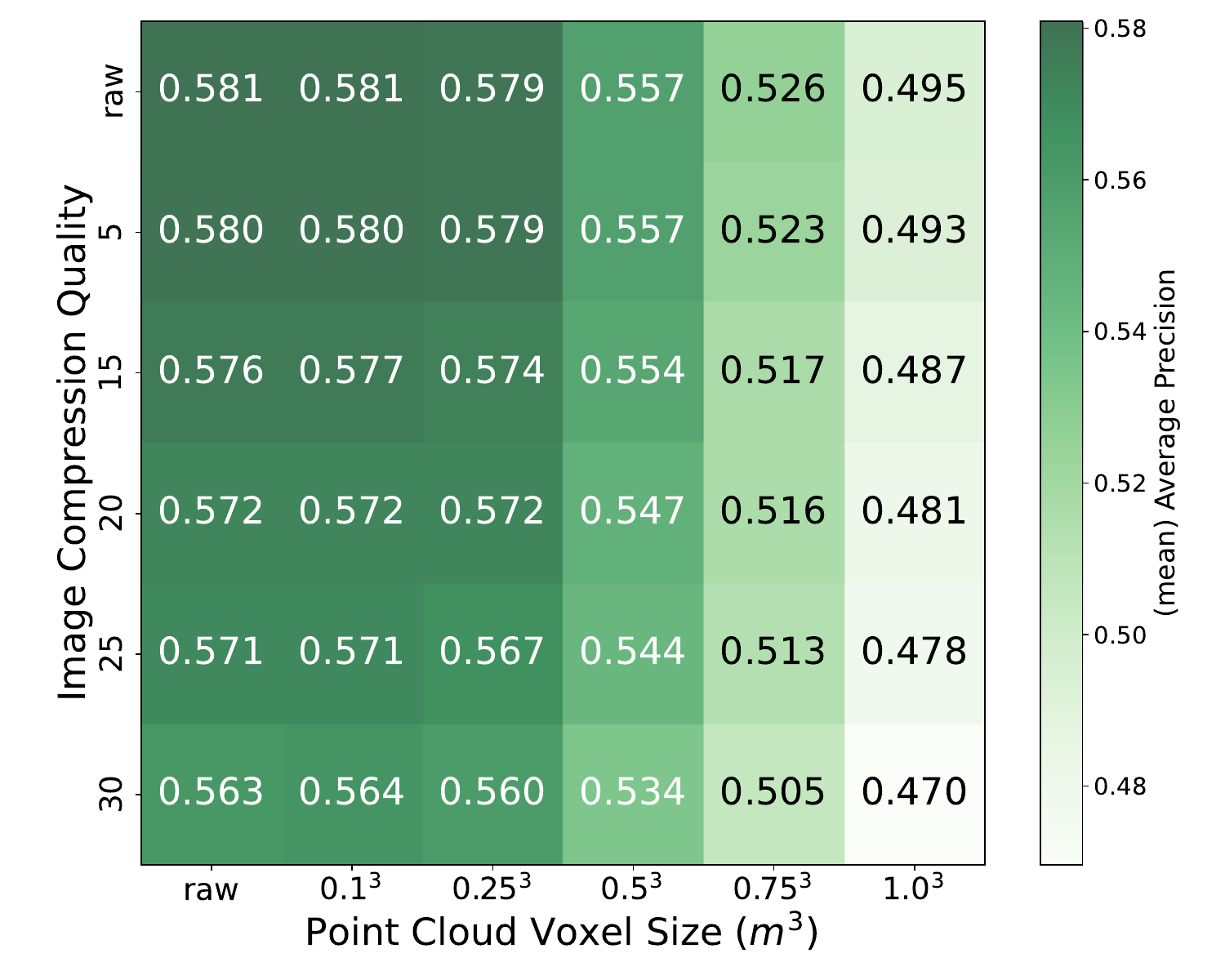}
    \caption{\small Heatmap of mAP as a function of point cloud voxel size and image compression quality. The higher precision values appear in the top-left region where compression quality is higher and voxel size is smaller.}
    \label{fig:compression_voxel_heatmap}
\end{minipage}
\end{figure*}
\subsection{Impact of Data Compression on Object Detection} 

\subsubsection{Camera-only object detection} 
We first evaluate the impact of compression on 3D object detection based only on the camera-captured images.
Fig \ref{fig:compression_quality} shows the impact of 

image compression quality factor on the Average Precision (AP) for different object classes when using only camera data. 
The AP for all classes decreases steadily as the compression quality factor increases from the original to 30. The overall mAP, represented by the red line, shows a significant drop, indicating that higher compression negatively impacts detection performance. This drop is particularly pronounced for “Pedestrian” and “Motorcycle,” indicating that the detailed features required to detect these objects are more likely to be lost due to compression.

\subsubsection{LiDAR-only object detection}
Fig \ref{fig:voxel_size} explores the effect of voxel size on the mean Average Precision (mAP) for various object classes using only LiDAR data. As voxel size increases from native to 1.0m$^3$, mAP for all classes drops significantly. The "traffic cone" and "pedestrian" classes are the most affected, with precision dropping dramatically as voxel size increases. As the size of the voxels increases, the AI's ability to detect smaller objects is significantly hampered due to the loss of detail. This emphasizes the crucial balance needed between maintaining data accuracy and compressing data to meet the limitations of 5G uplink.

In 3D object detection, LiDAR generally achieves a better mAP score compared to camera-based methods due to its superior depth perception and accuracy in spatial measurements. However, for specific classes such as "traffic cone", the impact of image compression is noticeably less severe than the impact of LiDAR point cloud compression. This indicates that while LiDAR provides better overall performance, its susceptibility to compression highlights the importance of maintaining high-quality data, especially for detecting smaller and more intricate objects where detail preservation is crucial.

\subsubsection{Multi-modal object detection}
Fig \ref{fig:compression_voxel_heatmap}
depicts a heat map showing the combined effect of point cloud voxel size and image compression quality on the average precision of object detection using fused information from both camera and LiDAR. The AP is highest in the upper left region of the heat map because the image compression quality is high (lower quality factor) and the voxel size is small in this region. In contrast, the lower right region has lower compression quality and larger voxels, resulting in the lowest precision value. This heat map highlights the importance of optimizing voxel size and compression quality to achieve the best detection performance.

\begin{figure*}[t] 
\begin{minipage}{.3\linewidth}
    \centering
    \includegraphics[width=\textwidth, height=1.7in]{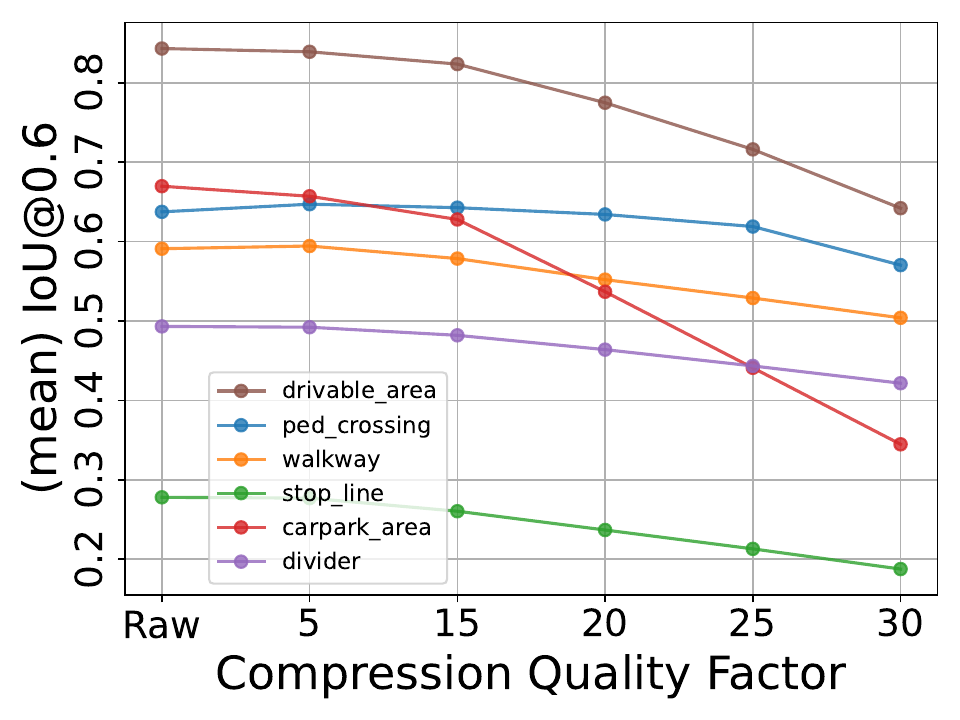} 
    \caption{Impact of compression quality factor on IoU@0.6 for various map segmentation categories. }
    \vspace{-4mm}
    \label{fig:compression_quality_segmentation}
\end{minipage}
\hfill
\begin{minipage}{.3\linewidth}
    \includegraphics[width=\textwidth, height=1.7in]{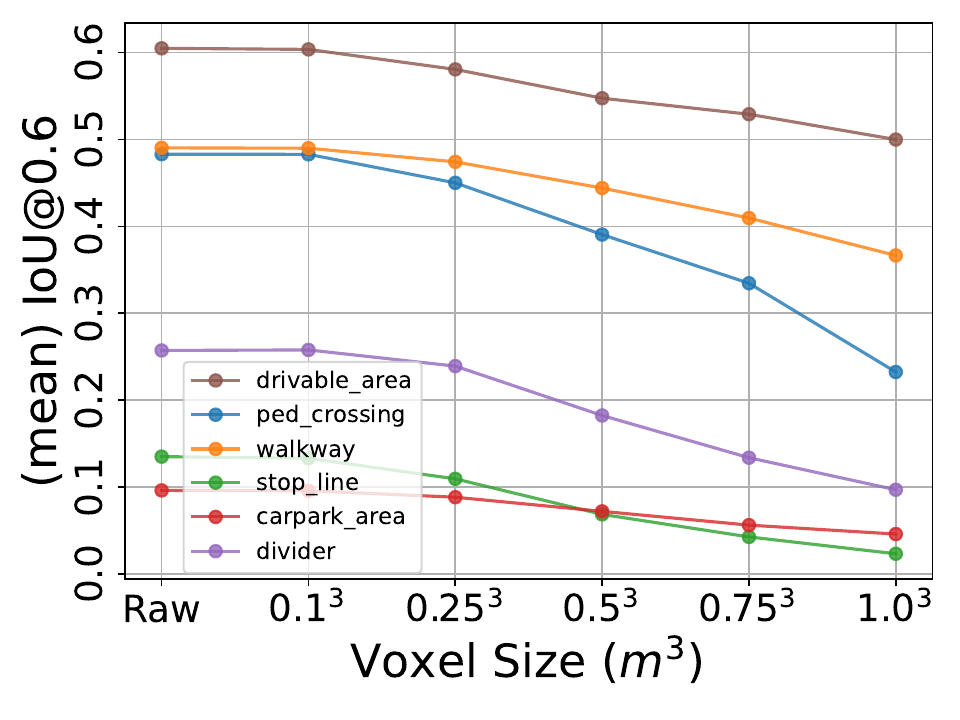}
    \caption{Impact of voxel size on (mean) IoU@0.6 for different map segmentation categories. }
    \vspace{-4mm}
    \label{fig:voxel_size_segmentation}
\end{minipage}
\hfill
\begin{minipage}{.37\linewidth}
    \centering
    \includegraphics[width=0.85\textwidth]{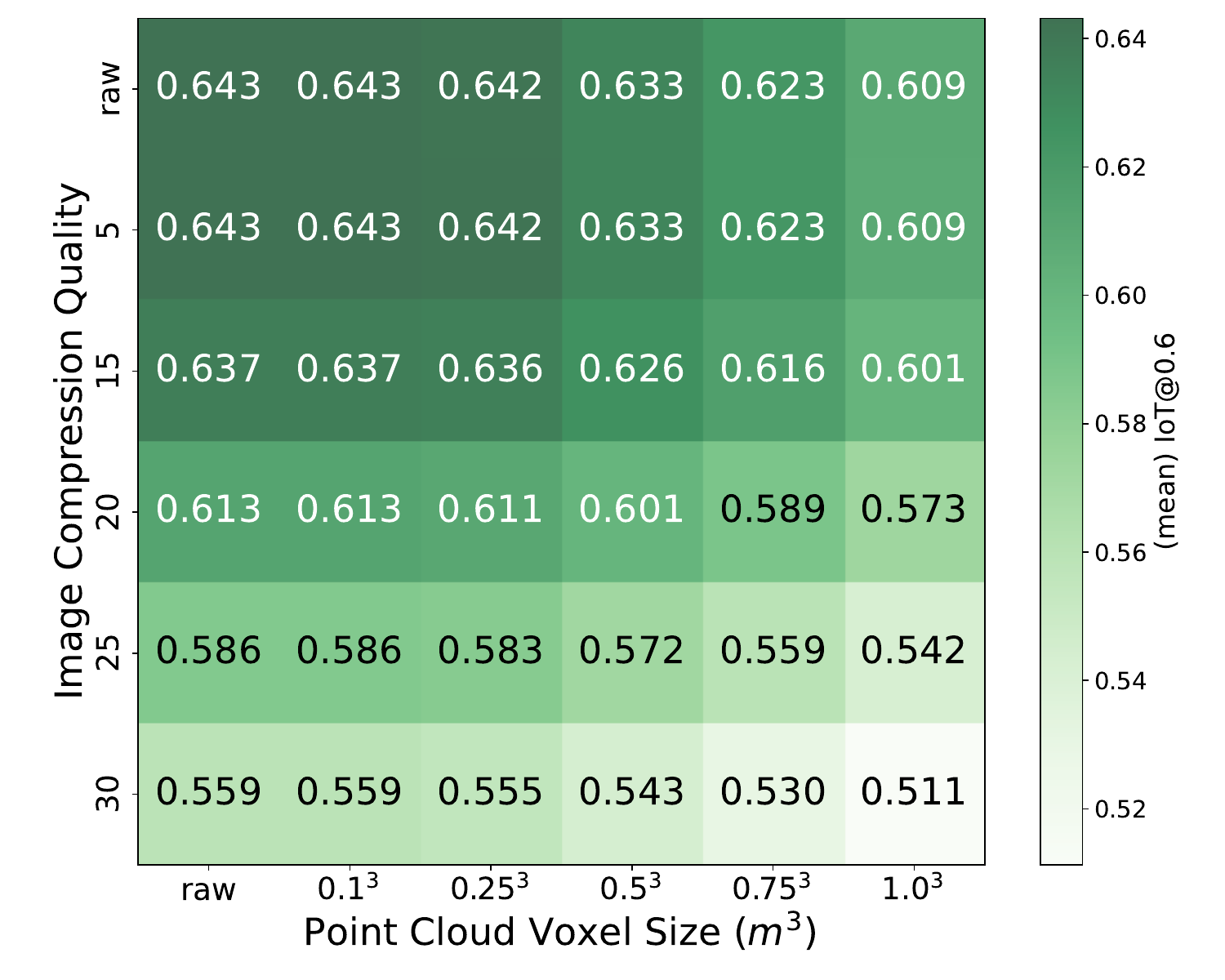}
    \caption{\small Heatmap of (mean) IoU@0.6 as a function of point cloud voxel size and image compression quality for map segmentation in BEVFusion. The higher IoU values appear in the top-left region where compression quality is higher and voxel size is smaller.}
    \vspace{-4mm}
    \label{fig:compression_voxel_heatmap_segmentation}
\end{minipage}
\end{figure*}

\subsection{Impact of Data Compression on  Segmentation} 

\subsubsection{Camera-only Segmentation}
Fig \ref{fig:compression_quality_segmentation}
illustrates the effect of varying the image compression quality factor on the average Intersection over Union (IoU) for different map segmentation classes in BEVFusion, with a threshold of 0.6. As the compression quality factor increases from the original (no compression) to 30, the IoU for all classes decreases significantly. The performance of the 'drivable\_area' and 'ped\_crossing' classes decreases significantly, indicating that higher compression rates have a detrimental effect on the accuracy of these segmentations. This trend suggests that lower compression quality (which preserves more image details) is critical for the accurate segmentation of these specific classes.

\subsubsection{Lidar-only Segmentation}
Fig \ref{fig:voxel_size_segmentation}
presents the effect of voxel size on the average IoU at a threshold of 0.6 for various map segmentation classes in BEVFusion when using LiDAR data. As the voxel size increases from the original (no voxelization) to 1.0 cubic meters, the IoU for all classes decreases significantly, especially for ``divider" and ``stop line". This decrease highlights the importance of finer voxel resolution for capturing detailed structure and maintaining high segmentation accuracy. Larger voxel sizes result in a loss of spatial resolution, which negatively affects the accuracy of smaller and more detailed segmentation tasks.
Therefore, having a high spatial resolution is crucial for maintaining accurate segmentation, especially when dealing with compression limitations due to network performance. The relationship between compression, uplink bandwidth, and AI accuracy in object detection and segmentation highlights the importance of optimizing these factors together. This leads us to explore how using multi-modal approaches, combining different data types, can help address these challenges.

\subsubsection{Multi-modal object Segmentation}
Fig \ref{fig:compression_voxel_heatmap_segmentation}
combines the effect of point cloud voxel size and image compression quality on the average IoU for a map segmentation threshold of 0.6 in BEVFusion. The heatmap shows that the highest IoU values are achieved with high image compression quality (lower quality factor) and small voxel size. Conversely, low compression quality combined with a larger voxel size leads to the lowest IoU values. The figure highlights the necessity of optimizing these two parameters to achieve the best segmentation performance, with the upper left area of the heatmap representing the best combination of high compression quality and small voxel size.

\begin{figure}[t]
\begin{minipage}{\linewidth}
    \centering
    \includegraphics[width=0.7\textwidth]{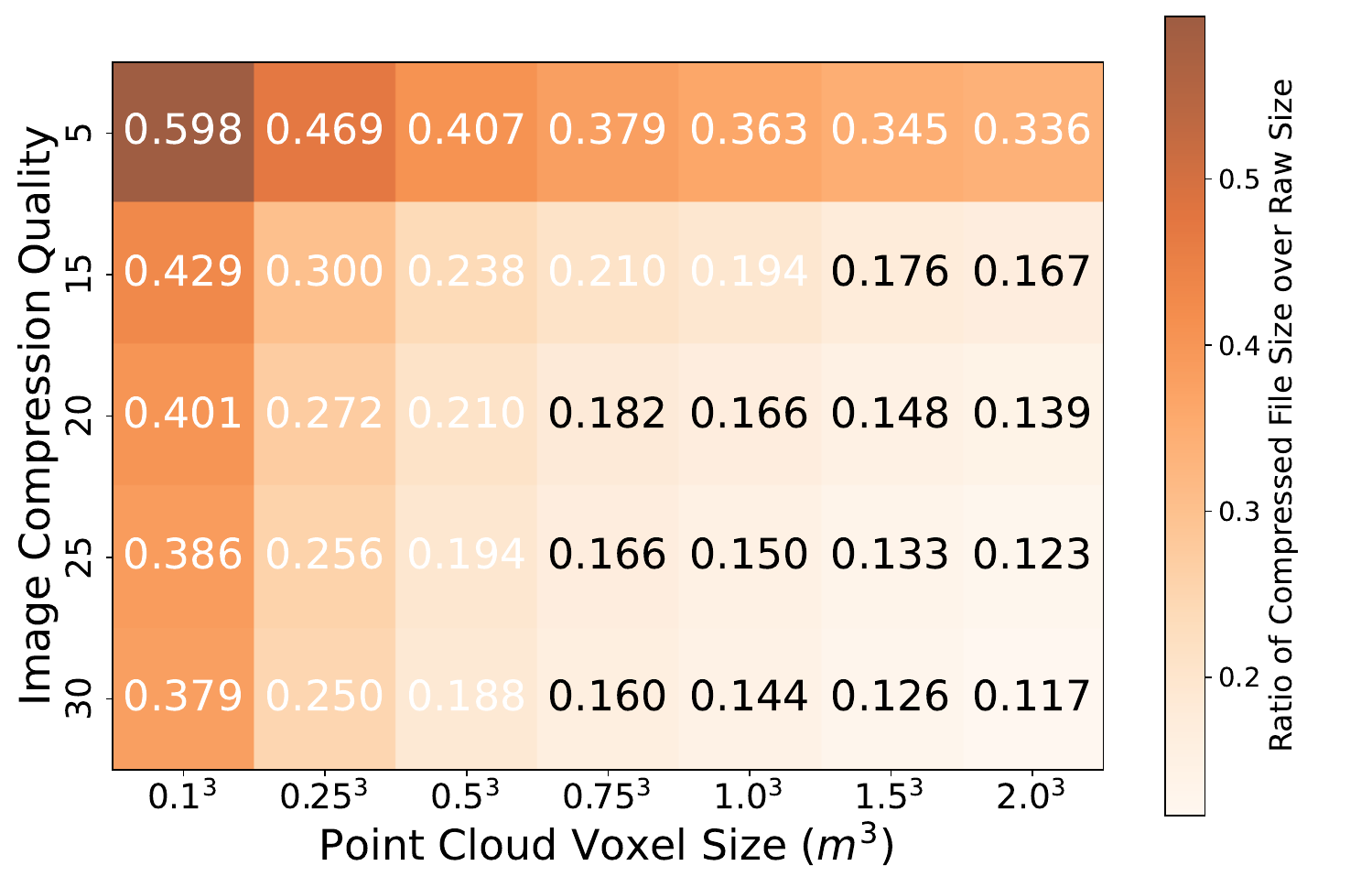}
    \caption{Ratio of compressed file size over raw file size for varying image compression quality and point cloud voxel size. The color indicates the ratio, with darker colors representing higher ratios.}
    \label{fig:ratio_compressed_filesize}
\end{minipage}
\end{figure}

\begin{figure}[t]
\begin{minipage}{\linewidth}
    \centering
    \includegraphics[width=0.7 \textwidth]{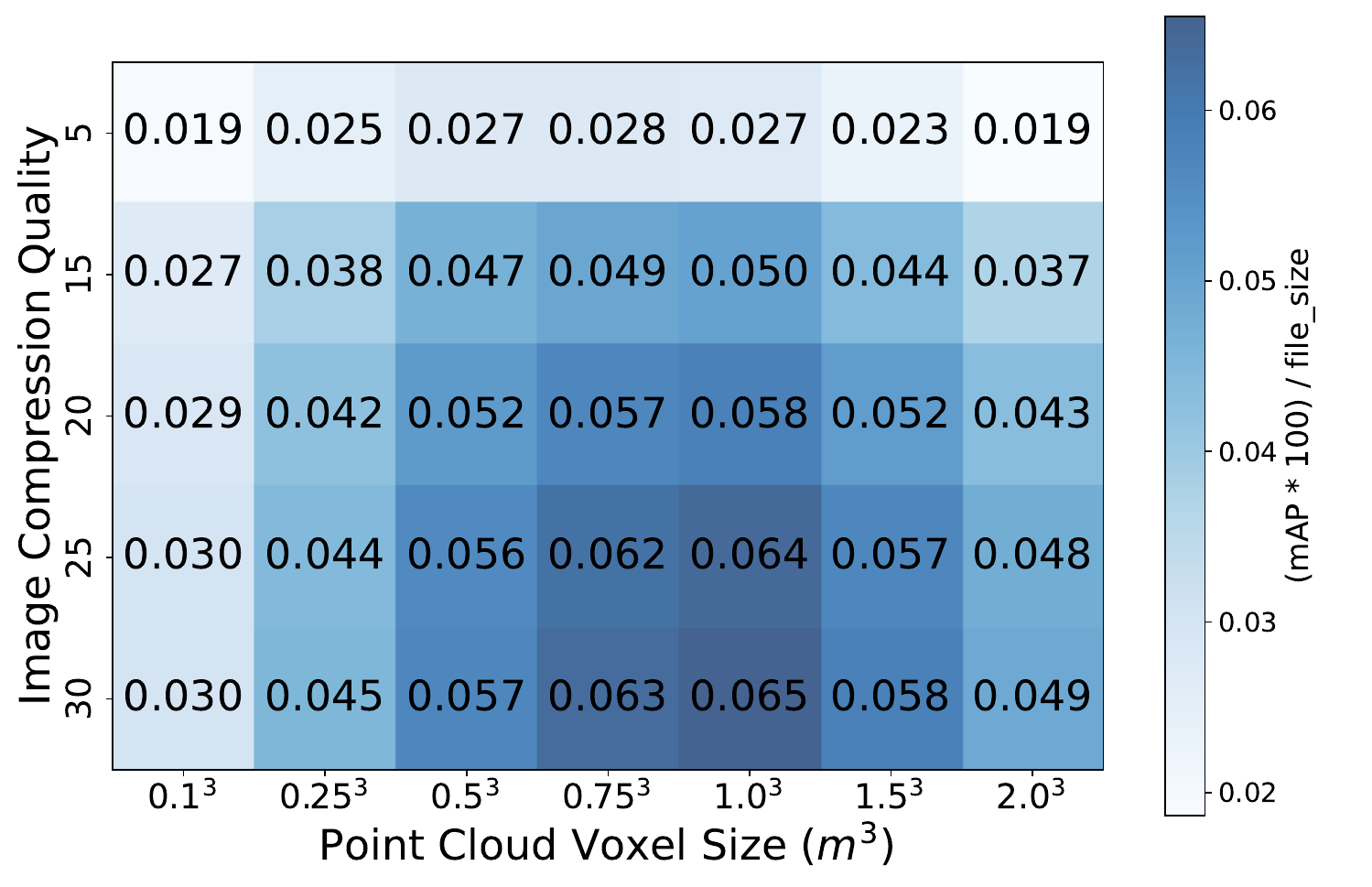}
    \caption{\small Impact of image compression quality and point cloud voxel size on detection performance. This heatmap shows the mean average precision (mAP) of object detection, with the x-axis representing point cloud voxel size (m³) and the y-axis representing image compression quality. Darker shades indicate higher mAP values. }
    \vspace{-4mm}
    \label{fig:compression_vs_voxel}
\end{minipage}
\end{figure}
\subsection{Compression and 3D Object Detection Efficiency}
Fig \ref{fig:ratio_compressed_filesize} and fig \ref{fig:compression_vs_voxel} provide a comprehensive view of the trade-offs between file compression and detection performance in terms of storage efficiency and accuracy. The first figure shows the ratio of compressed file size over raw file size for various image compression qualities and point cloud voxel sizes, where darker colors indicate higher compression ratios. The second figure displays the ratio of mAP multiplied by 100 over file size for the same range of parameters, with darker colors representing higher detection performance relative to file size. 
From Fig \ref{fig:ratio_compressed_filesize}, we observe that as the point cloud voxel size increases, the file compression ratio decreases, indicating that larger voxel sizes result in smaller compressed files relative to the raw file size. This trend is consistent across different image compression qualities, although the highest compression ratios are seen at lower voxel sizes and lower image compression qualities.

Fig \ref{fig:compression_vs_voxel} illustrates the relationship between the image compression quality, point cloud voxel size, and the ratio of mAP multiplied by 100 over file size. The color intensity in the heatmap represents this ratio, where darker shades indicate higher values. From the analysis, it is evident that higher point cloud voxel sizes generally lead to better performance in terms of the mAP-to-file-size ratio. Specifically, the highest ratios are observed at a voxel size of $0.75^3 \, \text{m}^3$ and an image compression quality of 30, suggesting an optimal balance between file size and performance. As the voxel size increases beyond this point, the ratio slightly decreases, indicating diminishing returns. Additionally, lower image compression qualities tend to have lower ratios, highlighting the trade-off between file size and detection performance. This analysis highlights the significance of optimizing image and point cloud compression together to enhance detection performance and reduce data size. By addressing object detection and segmentation for single and multi-modal tasks, we show that the limitations of 5G uplink capacity play a crucial role in determining the efficiency of AI-driven teleoperation.

\begin{figure}[t] 
\begin{minipage}{.48\linewidth}
    \centering
    \includegraphics[width=\textwidth, height=1.2in]{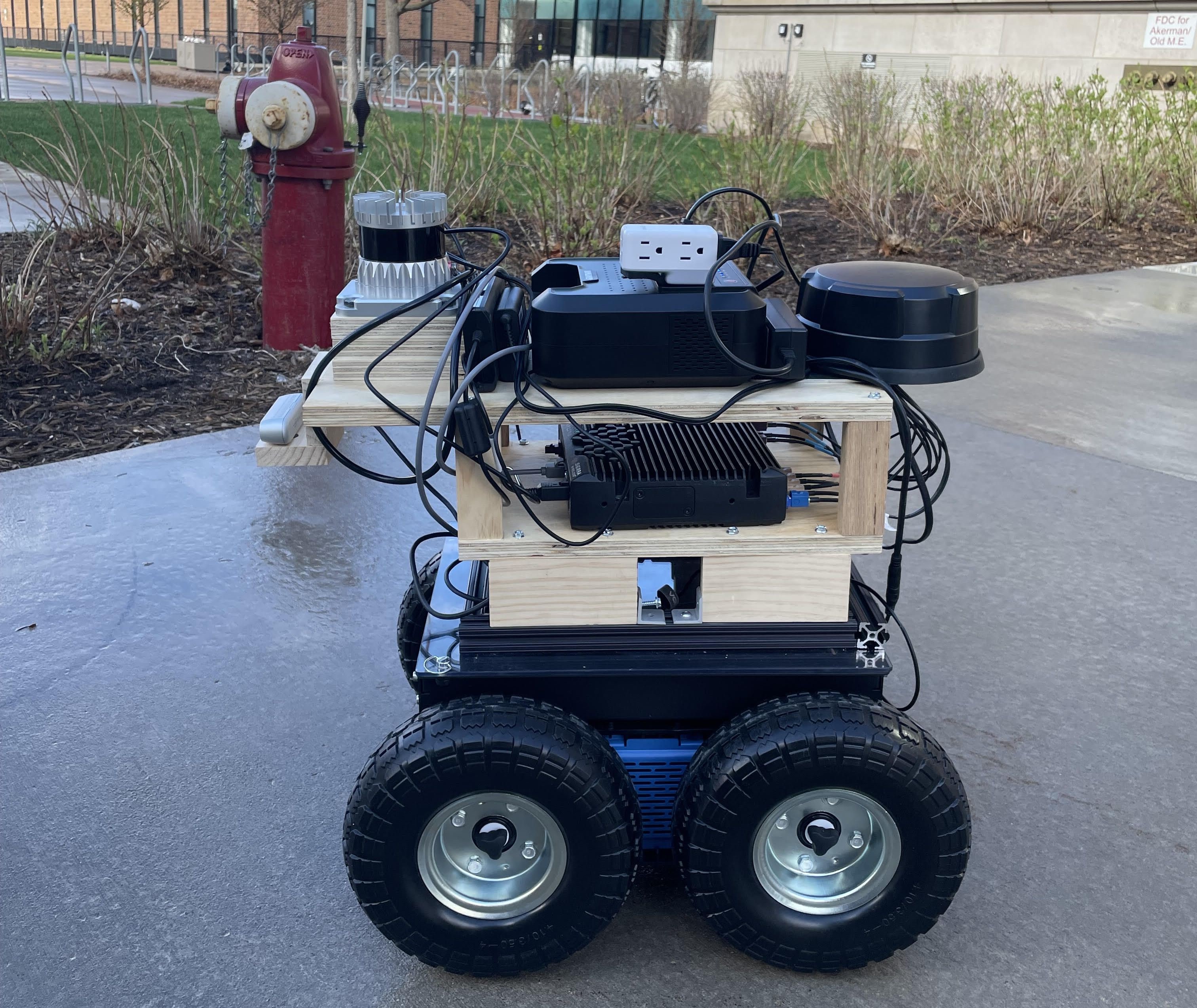} 
    \caption{Robotic vehicle.} 
    \vspace{-4mm}
    \label{fig:UMN_AV}
\end{minipage}
\hfill
\begin{minipage}{.48\linewidth}
    \includegraphics[width=\textwidth, height=1.2in]{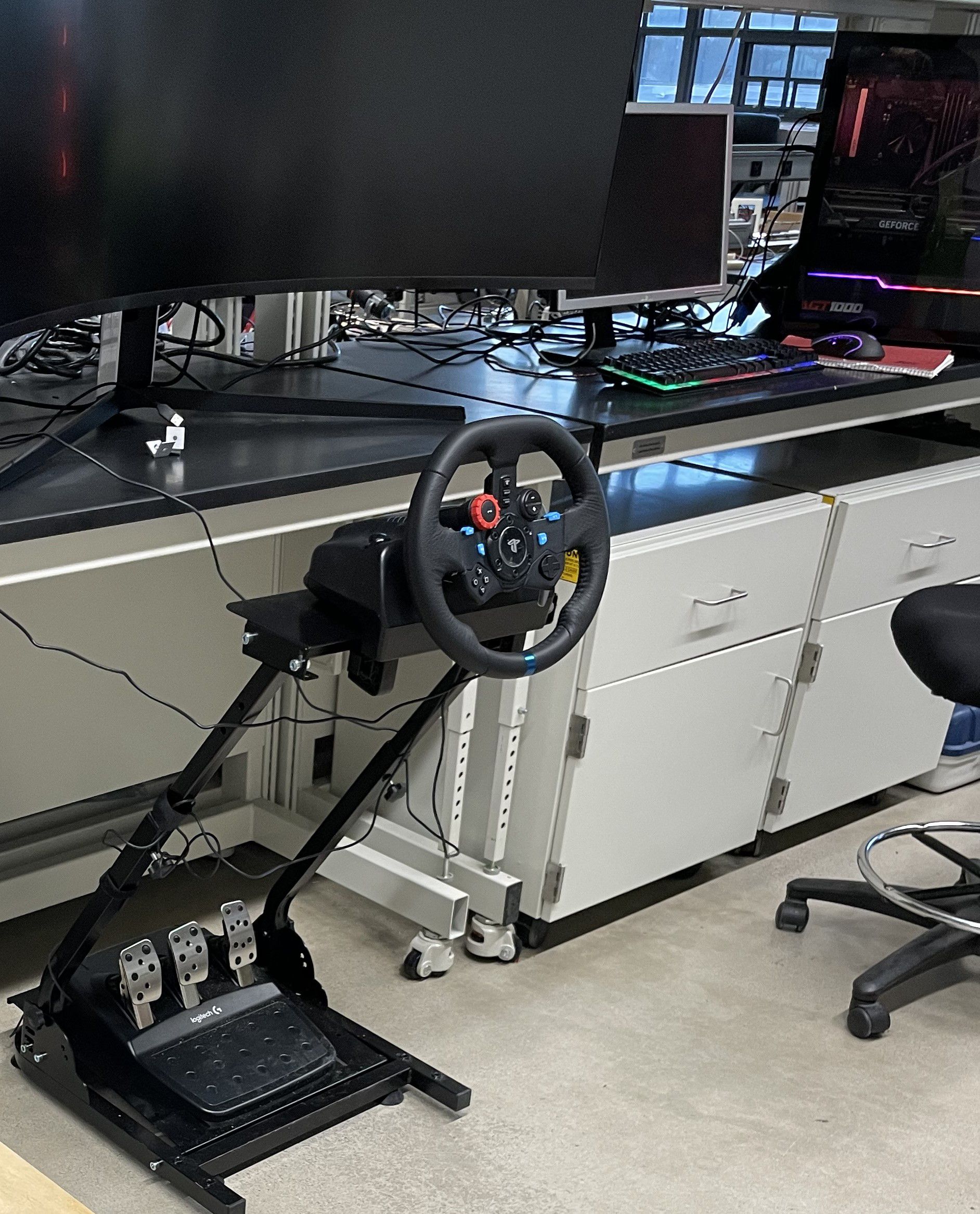}
    \caption{Teleoperation cabin.} 
    \vspace{-4mm}
    \label{fig:UMN_teleoperation}
\end{minipage}
\end{figure}

\section{Conclusion}
In summary, this paper highlights the crucial role of data compression in enabling reliable teleoperation over 5G networks, where uplink limitations demand careful data management. Our research indicates that using lossy compression can reduce the performance of AI. We also suggest methods to balance data fidelity with network limitations to optimize teleoperation. Future work will explore additional AI algorithms, refine compression methods, and test real-time transmission using our current autonomous vehicle setup, as shown in Fig.~\ref{fig:UMN_AV} and \ref{fig:UMN_teleoperation}.

\section{Acknowledgment}
We thank the anonymous reviewers for their suggestions and feedback. This research is supported in part by the National Science Foundation (NSF) under grants  2212318, 2220286, 2220292  and 2321531 as well as an InterDigital gift.

\bibliographystyle{IEEEtran}
\bibliography{bib/refs.bib}

\end{document}